\documentclass[12pt]{article}
\usepackage{graphicx}
\usepackage{color}
\usepackage{amsmath}
\usepackage{amssymb}
\usepackage{subcaption}
\usepackage[margin=1in,footskip=0.25in]{geometry}
\providecommand{\keywords}[1]
{
  \small	
  \textbf{\textit{Keywords---}} #1
}
\begin{document}

%%%%%%%%%%%%%%%%%%%%%%%%%%%%%%%%%%%%%%%%%%%%%%%%%%%%%%%%%%%%%%%%%%%%%%%%%%%%%%

  \title{\bf  Dynamics of Chemical Orders in Formation of Striped  Patterns in Metamorphic Rocks.}
  \author{Bikash Kumar Sarkar \thanks{Department of Physics, Mrinalini Datta Mahavidyapith, Birati,
Kolkata, 700 051, West Bengal, India.} \\
    Swayambhoo Mitra \thanks{Department of Physics, Barrackpore Rastraguru Surendranath College,
85, Middle Road, 6, River Side Rd, Kolkata, 700120, West Bengal, India.} \\
Manas Kumar Roy \thanks{Department of Physics, Brahmananda Keshab Chandra College, 111/2
B.T. Road, Bonhooghly, Kolkata, 700108, West Bengal, India.}\\
Biswajit Saha \thanks{
    Department of Physics, Gobardanga Hindu College, Gobardanga,
743273, West Bengal, India}\hspace{.2cm}}
\maketitle
\begin{abstract}  
The striped patterns in rocks, characterized by thin dark bands and thick light bands, are a result of the natural processes that have shaped the rocks through in situ metamorphism. Notably, the dark and light bands are composed of similar minerals, with the dark bands containing additional trapped materials such as fluid pockets or extra mineral grains. Here we employed the Gray-Scott Reaction Diffusion model to investigate the interaction between two sets of virtual chemicals, denoted as `u' and `v'. 
The differing diffusion and reaction rates of `u' and `v' chemical orders lead to the formation of $180^{\circ} $ out-of-phase chemical domains, resulting in striped patterns. Utilizing the Gray-Scott model in this manner, we gain valuable insights into the early microscopic stages of these geologically significant striped patterns in metamorphic rocks.
\end{abstract}
\section{Introduction}

Striped patterns in rocks are visually striking and mysterious textures characterized by alternating dark and bright bands, ranging from millimeter to centimetre scales, found within sedimentary layers. 
These textures are called by different names such as `diagenetic crystallization rhythmites' \cite{fontbote1980, fontbote1981, fontbote1982}, expansion structures \cite{beales1980}, rhythmites \cite{sass1994}, banded or ribbon ores \cite{sass1982, tompkins1994},  or more commonly as zebra rock, fabric, texture, dolomite \cite{beales1980, wallace1994, nielsen1998,   morrow2014}.
striped textures are often composed of dolomite, although similar textures are also found in minerals such as ankerite, siderite, sphalerite, marcasite, barite, fluorite, magnesite, and others \cite{sass1994, wallace1994, arne1989, fontbote1993}.
These banded textures hold significant economic importance as they frequently occur with carbonate-hosted zinc-lead mineralization \cite{sass1994, tompkins1994, arne1989,  cordeiro2018}, often forming the predominant ore textures in such deposits. striped textures are also observed in carbonate reservoir rocks, especially in `hydrothermal dolomite reservoirs' \cite{mccormick2023,
davies2006, hiemstra2015}. While earlier researchers favored a synsedimentary or early diagenetic origin for these structures \cite{fontbote1980, fontbote1981, fontbote1982}, more recent studies lean towards a late diagenetic origin.The banding in zebra textures is caused by various processes, including the replacement of sedimentary structures like evaporite laminae \cite{beales1980, tompkins1994}, dolomite replacement \cite{sass1982, arne1989}, fracturing \cite{wallace1994, nielsen1998, lopezhorgue2009}, dissolution \cite{morrow2014}, hydraulic over-pressuring \cite{swennen2003, boni2000,  vandeginste2005, swennen2012}, and grain growth during recrystallization \cite{kelka2015, kelka2017}. Dolomitization occurs when CaCO\(_3\) is replaced by CaMg(CO\(_3\))\(_2\) through crystal-scale dissolution-precipitation, as described by \cite{putnis2002}. Dolomite cementation, on the other hand, involves the direct precipitation of CaMg(CO\(_3\))\(_2\) from an aqueous solution, as defined by \cite{machel2004}. Dolomite recrystallization is a broader term encompassing both low-temperature diagenetic recrystallization, where existing crystals are replaced by new, thermodynamically stable ones of the same mineralogy, and high-temperature dynamic recrystallization, which involves grain boundary migration, as explained by \cite{machel1997} and \cite{newman1994} respectively. Scientists are intrigued by the unique appearance of striped patterns in rocks and the complex mechanisms behind their formation. Various hypotheses have been proposed to explain the exact processes leading to these patterns \cite{wallace2018}. Despite extensive research, the initial stages of striped pattern formation remain elusive. 

In this study, we use the Gray-Scott reaction-diffusion model to investigate how striped patterns form in rocks \cite{gray1984}. Our model illustrates how chemical reactions and movement can generate striped-like patterns from an uneven starting point. By adjusting model factors, we explore the conditions that lead to these patterns. Comparing our model's findings with real striped patterns helps validate our hypotheses and offers insights into the early processes that shaped these distinct rock patterns. Understanding the origins of striped patterns informs our knowledge of mineral formation and aids in identifying potential ore deposits.

\section{Striped Patterns and Gray-Scott Reaction-Diffusion Model}

The Gray-Scott model is a mathematical representation of a non-linear reaction-diffusion system involving two chemical species, denoted as $u$ and $v$. In this model, the reactions occur as follows:

\begin{eqnarray}
    u + 2v &\rightarrow & 3v\\
    v &\rightarrow & P
\end{eqnarray}

Here, $u$ is continuously introduced into the system, while the inert product $P$ is removed\cite{buric2014}. The model is based on the law of mass action\cite{har2016}, which assumes that the rate of each reaction is proportional to the concentration of the reactants at each point in the system. The Gray-Scott reaction-diffusion model consists of two coupled partial differential equations that describe the evolution of two chemical species, `u' and `v', over time. The equations for `u' and `v' are given by:

\begin{equation}
\frac{\partial u}{\partial t} = D_u \nabla^2 u - uv^2 + F(1 - u)=F_1(u,v)
\label{eq1}
\end{equation}

\begin{equation}
\frac{\partial v}{\partial t} = D_v \nabla^2 v + uv^2 - (F + K)v =F_2(u,v)
\label{eq2}
\end{equation}

Here, $u = u(t, x)$ and $v = v(t, x)$ represent the (dimensionless) concentrations of the chemical species $u$ and $v$, respectively. The Laplacian operator $\nabla^2$ with respect to the spatial variable $x$ is used, and $D_u$ and $D_v$ are the diffusion coefficients of $u$ and $v$, respectively. The parameter $F$ represents the rate of $u$ feed and removal of $u$, $v$, and $P$ from the system, while $K$ is the rate of conversion of $v$ to $P$. The diffusion coefficients \( Du \) and \( Dv \) are typically considered constant, with \( D_u / D_v = 2 \).

The core equations for the model are as follows:

\begin{equation}
u_{i,j}^{n+1} = u_{i,j}^{n} + \Delta t \left( D_u \nabla^2 u_{i,j}^{n} - u_{i,j}^{n} (v_{i,j}^{n})^2 + F (1 - u_{i,j}^{n}) \right)
\end{equation}

\begin{equation}
v_{i,j}^{n+1} = v_{i,j}^{n} + \Delta t \left( D_v \nabla^2 v_{i,j}^{n} + u_{i,j}^{n} (v_{i,j}^{n})^2 - (F + K) v_{i,j}^{n} \right)
\end{equation}

Here, $u^{n}_{i,j}$ and $v^{n}_{i,j}$ represent the concentrations of `u' and `v' at grid point $(i, j)$ and time step `n'. The Laplacian operator $\nabla^2$ is approximated using finite differences in terms of neighboring grid points. Here's an interpretation of each term:
\begin{enumerate}
   \item \textbf{Rate of Change Terms:}
   $\frac{\partial u}{\partial t}$ and $\frac{\partial v}{\partial t}$ represent the rates of change of the concentration of `u' and `v' over time, respectively.
   
   \item \textbf{Diffusion Coefficients:}
   $D_u$ and $D_v$ are the diffusion coefficients for `u' and `v', respectively. They control the rate at which `u' and `v' diffuse in space. Larger values of $D$ correspond to faster diffusion.
   
   \item \textbf{Spatial Variation Terms:}
   $\nabla^2 u$ and $\nabla^2 v$ represent the Laplacian of `u' and `v'. The Laplacian quantifies how concentrations change in space and is computed as the sum of second spatial derivatives. 
   
   \item \textbf{Reaction Terms:}
   $u v^2$ represents the reaction between `u' and `v'. They describe the production of `u' and the consumption of `v' through the reaction process.
   \item \textbf{Production and Decay Rates:} $F$ and $(F + K)$ represent the production rates of `u' and the effective production rate of `v' after considering the decay rate $K$. 

   \item \textbf{Grid and Time Parameters:} $n$ represents the number of grid points, which discretize the spatial domain into an \(n \times n\) grid. \(dx\) and \(dy\) represent the spatial steps in the \(x\) and \(y\) directions on the grid, calculated as \(1/n\) in each direction. \(dt\) is the time step, determining the size of each time increment in the numerical simulation.
   \item \textbf{Initial Conditions}
   The initial concentrations of `u' and `v' at time \(t = 0\) are defined as: The initial concentration of `u' is given by:
\begin{equation}
\ u(0, x) = 1 - e^{-80(x + 0.05)} \
\end{equation}
In this equation, \(x\) represents the spatial coordinate, and the initial concentration of `u' is determined for different positions on the spatial grid. The term (x + 0.05 ) shifts the spatial coordinate by 0.05 units, altering the initial concentration profile of `u' along the spatial domain.
The initial concentration of `v' is given by:
\begin{equation}
\ v(0, x) = e^{-80x} \
\end{equation}

Similar to `u', \(x\) represents the spatial coordinate. The expression \(e^{-80x}\) describes the initial concentration profile for `v', which starts close to 1 and exponentially decreases as \(x\) increases. These initial conditions establish the initial distribution of concentrations for `u' and `v' at the beginning of the simulation, capturing how they are spread across the spatial domain in the context of the Gray-Scott reaction-diffusion model. The chosen initial profiles of `u' and `v' set the stage for stripe formation. The spatial shift in `u' and the exponential decrease in `v' play a crucial role in shaping the striped Pattern.

\item \textbf{Boundary Conditions} Boundary conditions define how $u$ and $v$ behave at the edges of your spatial domain. Common boundary conditions include periodic (values wrap around the domain), no-flux (zero gradient), or Dirichlet conditions (fixed values).
\begin{itemize}

\item \textbf{Dirichlet Boundary Conditions} Dirichlet conditions involve specifying fixed values for $u$ and $v$ at the domain boundaries. In this code, we apply Dirichlet conditions on the left and right boundaries for $u$ and $v$. The left boundary of $u$ is set to $0.5$, and the right boundary of $v$ is set to $0.2$.

\begin{verbatim}
Define Dirichlet boundary conditions
(fixed values)
left_boundary_u = 0.5;  
right_boundary_v = 0.2;  
\end{verbatim}

Within the time loop, we update the concentration fields while respecting these Dirichlet conditions:
\begin{verbatim}
% Update u using vectorized operations
u = u + dt * (Du * del2(u, dx, dy) 
+ F * (1 - u) - u .* v.^2);
v = v + dt * (Dv * del2(v, dx, dy) 
- K * v + u .* v.^2);

% Apply Dirichlet boundary conditions
u(1, :) = left_boundary_u; 
% Left boundary (Dirichlet)
u(n, :) = right_boundary_v;
% Right boundary (Dirichlet)
\end{verbatim}

\item \textbf{Neumann Boundary Conditions}
Neumann conditions enforce a zero gradient (no-flux) boundary, meaning there is no net flow of the substances at the boundaries. In this code, Neumann conditions are applied to the top and bottom boundaries for both $u$ and $v$ by setting the concentration gradients at these boundaries to zero. This ensures that there is no diffusion across these boundaries.

\begin{verbatim}
% Apply Neumann boundary conditions 
(zero gradient)
u(:, 1) = u(:, 2);  
% Bottom boundary (Neumann)
u(:, n) = u(:, n - 1);  
% Top boundary (Neumann)
v(:, 1) = v(:, 2);  
% Bottom boundary (Neumann)
v(:, n) = v(:, n - 1);  
% Top boundary (Neumann)
\end{verbatim}
\end{itemize}
\end{enumerate}

\section{Stability Analysis} Regarding the stability analysis of the Gray-Scott system, which describes the interaction between chemical species $u$ and $v$ through their concentrations $u$ and $v$ in the time-space domain, the equations demonstrate the rate at which these concentrations change over time. The right-hand side of each equation comprises three terms: the reaction term, proportional to the product of the concentrations of $u$ and $v$ squared ($uv^2$), represents the rate of the reaction that consumes $u$ and generates $v$. The parameter $F$, known as the feed rate, controls the replenishment of $u$ through the term $F(1-u)$, while $Du$ represents the diffusion of $u$. Similarly, the second equation includes the term $(F + K)v$, where $K$, termed the kill rate, regulates the conversion of $v$ to an inert product P and is multiplied by $v$ due to $v$'s concentration dependency.

Stability analysis is crucial in understanding physical systems governed by partial differential equations, especially regarding the existence of equilibrium solutions and their dependence on the problem's parameters. For a physical problem to be well-posed, including the existence, uniqueness, and continuous dependence of the solution on the data, stability of the solution must be checked. In this context, we focus on the stability of the uniform steady state of the Gray-Scott model, which determines the system's evolution over time after reaching a certain threshold. We employ a theorem to analyze this stability aspect.

To apply this model to the formation of striped rock patterns, we consider the initial concentration distribution of one of the chemical species (let's say $u$) to be higher in the dark layers of the rock due to the presence of inclusions. This distribution is influenced by sedimentary processes and local geological conditions. As the Gray-Scott model equations evolve over time, the chemical species $u$ and $v$ interact and diffuse through the rock matrix. The differential diffusion rates and reaction kinetics of these chemicals could lead to the formation of localized patterns, resembling striped patterns.

Through numerical simulations and sensitivity analyses of the Gray-Scott model parameters, we aim to understand the conditions under which striped patterns may emerge in rocks. By aligning the model's predictions with observed striped patterns in real rocks, we can validate the viability of this microdynamic approach to explain the initial stages of pattern formation. This study bridges the gap between geological observations and microscale processes, offering insights into how complex geological features like striped patterns can emerge from chemical reactions and diffusion dynamics.
\subsection{Jacobian}

The Gray-Scott reaction-diffusion model comprises a pair of coupled partial differential equations governing the temporal evolution of two chemical species, \( u \) and \( v \). The equations (1) and (2) representing the dynamics of \( u \) and \( v \). In the domain of banded patterns, stability analysis plays a pivotal role, particularly when delving into the fascinating dynamics that underlie these distinctive patterns. Stability analysis is the process of scrutinizing the intricate behavior of the system and unraveling the conditions that govern the persistence of these captivating banded structures. It provides crucial insights into pattern stability and offers a glimpse into how these patterns evolve. One common method for conducting stability analysis is to employ linear perturbation analysis. This approach involves assuming small perturbations around a stable, uniform solution and subsequently examining the linearized versions of the governing equations. Understanding the system's stability hinges on a careful evaluation of the eigenvalues derived from these linearized equations. The Gray-Scott model, which revolves around a system of reaction-diffusion equations, holds significant relevance in the context of banded patterns found within hornfels. These equations serve as a mathematical representation of the dynamic evolution of mineral concentrations over time, granting us access to the intricate narrative of hornfels' mineralogical transformation and the emergence of these distinctive banded patterns. Two primary substances are considered:
\begin{enumerate}
\item {\(u\)} represents one of the key mineral constituents within the rock.
\item{\(v\) }represents another mineral constituent.
\end{enumerate}
Using Equations \ref{eq1} and \ref{eq2}

\[
\frac{\partial F_1}{\partial u} = - v^2 - F, \quad \frac{\partial F_1}{\partial v} = -2uv
\]

\[
\frac{\partial F_2}{\partial u} = v^2, \quad \frac{\partial F_2}{\partial v} =  2uv - (F + K)
\]

Jacobian matrix (\( \mathbf{J} \)) for the given system of equations \ref{eq1} and \ref{eq2}

\begin{equation}
\mathbf{J} = \begin{bmatrix}
 - v^2 - F & -2uv \\
v^2 &   2uv - F - K
\end{bmatrix}
\end{equation}

  The characteristic equation \( \text{det}(\mathbf{J} - \lambda \mathbf{I}) = 0 \) is:

\[
\left|\begin{matrix}
(- v^2 - F) - \lambda & -2uv \\
v^2 &   (2uv - F - K) - \lambda
\end{matrix}\right| = 0
\]

The eigenvalues of the Jacobian matrix are:
\[
\lambda_1 = uv - \frac{K}{2}  - F - \frac{v^2}{2} - \frac{{A}}{2}
\]
\[
\lambda_2 = uv - \frac{K}{2}  - F - \frac{v^2}{2} +  \frac{{A}}{2}
\]

where \( A = \sqrt {K^2 - 4Ku v - 2K v^2 + 4u^2 v^2 - 4uv^3 + v^4} \)

The trace of the Jacobian matrix \( J \) is:
\[
\text{trace}(J) = - v^2 + 2uv - 2F - K
\]

The determinant of the Jacobian matrix \( J \) is:
\[
\det(J) = F^2 + Fv^2 - 2uvF + Kv^2
\]

After Numerical analysis, In the context of the Gray-Scott model, the eigenvalue \( -0.051014 + 0.005822i \) has a negative real part \( \text{Re}(\lambda) = -0.051014 \). This negative real part indicates that, for all wavenumber (k), the steady state corresponding to the variables \( u \), \( v \), \( F \), and \( K \) is stable. This stability condition suggests that the system, governed by parameters \( D_u \), \( D_v \), \( F \), and \( K \), along with the steady-state values \( u_s \) and \( v_s \), is unlikely to exhibit instabilities under these conditions.
\begin{figure}[h!]
    \centering \includegraphics[width=0.72\textwidth, height=7 cm]{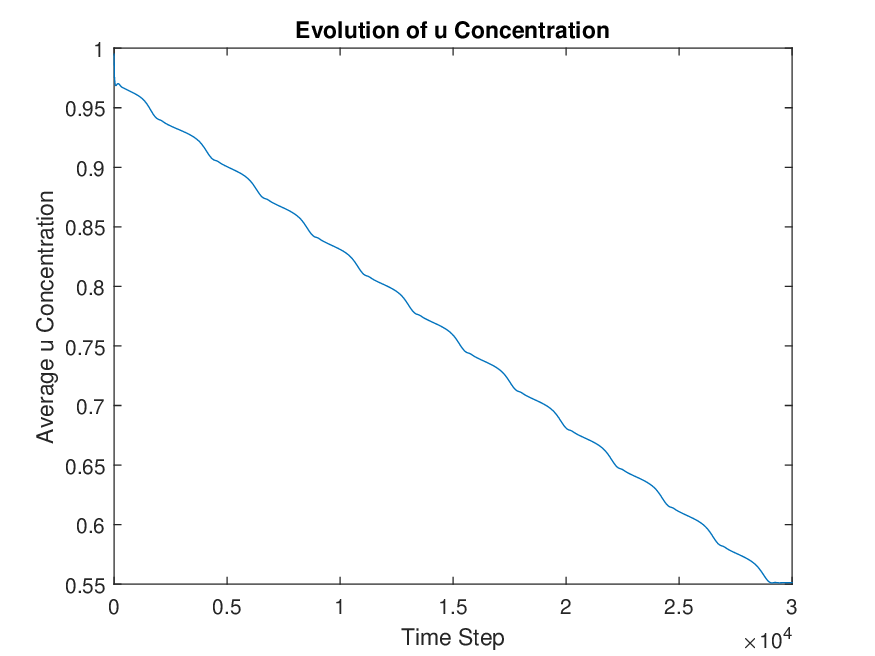}
    \caption{Stability}
    \label{fig:stability}
\end{figure}

\section{Formation Process of striped Rock}
\begin{itemize}
    \item \textbf{Pyrite Oxidation}: Pyrite oxidation initiates the release of Fe and sulfate ions into the environment\cite{kawahara2023}.
    
    \item \textbf{Acidic Fluid Infiltration}: Acidic fluids, resulting from the oxidation of pyrite, infiltrate surrounding sedimentary rocks\cite{kawahara2023}.
    
\item \textbf{Neutralization Reactions}: These acidic fluids react with sedimentary minerals, such as potassium–aluminum silicates (e.g., feldspars, mica), in neutralization reactions:
\[
\begin{aligned}
    & 3KAlSi_3O_8 + Fe^{3+} + 18H_2O + 2SO_4^{2-} + 3H^+ \\
    & \rightarrow KAl_3(SO_4)_2(OH)_6 + Fe(OH)_3 + 9H_4SiO_4 + 2K^+
\end{aligned}
\]
\[
\begin{aligned}
    & 2KAlSi_3O_8 + Fe^{3+} + 12H_2O \\
    & \rightarrow Al_2 Si_2 O_5(OH)_4 + Fe(OH)_3 + 4H_4SiO_4 + H^+ + 2K^+
\end{aligned}
\]

    These reactions precipitate Fe oxyhydroxide minerals and alter the sediment's mineral composition\cite{kawahara2023}.
    
    \item \textbf{Band and Spot Formation}: Fluctuations in pH buffering within sedimentary rocks lead to the formation of alternating bands of Fe oxyhydroxide and clay minerals. Concurrently, bleached spots develop through the dissolution of Fe minerals\cite{kawahara2023}.
\end{itemize}

This process is critical in understanding the formation mechanisms underlying the distinctive patterns observed in striped rock.

\section{Simulation Result}

For numerical simulation, we used the following parameters:
\[
\begin{aligned}
    & \text{Diffusion coefficient for } u: & Du &= 0.00002, \\
    & \text{Diffusion coefficient for } v: & Dv &= 0.00001, \\
    & \text{Production rate of } u: & f &= 0.04, \\
    & \text{Decay rate of } v: & k &= 0.1, \\
    & \text{Total number of time steps:} & tf &= 30000, \\
    & \text{Number of grid points:} & n &= 400, \\
    & \text{Time step:} & dt &= 0.5.
\end{aligned}
\]

In this section, we will compare the striped rock patterns generated through the Gray-Scott Reaction-Diffusion model in our numerical simulations with natural striped rock patterns observed in geological formations. By examining the similarities and differences, we can gain insights into the ability of our model to mimic real-world geological processes.

\subsection{Simulated striped Rock Pattern}

The striped rock pattern generated by our numerical simulation is a result of the Gray-Scott Reaction-Diffusion model in action. Through this simulation, we can visualize the evolving patterns of two reacting substances, \( u \) and \( v \), over time and space. The spatial distribution and temporal evolution of these substances give rise to intricate striped-like patterns. We have captured the simulated striped rock pattern at different time intervals to observe its development and complexity. These snapshots, represented as images, provide a visual representation of how the pattern forms and evolves over time. We have captured the simulated striped rock pattern at the following time intervals: (1, 1000, 5000, 10000, 15000, 20000, 25000, 30000) time steps.

\begin{figure}[h!]
    \centering \includegraphics[width=0.72\textwidth, height=7 cm]{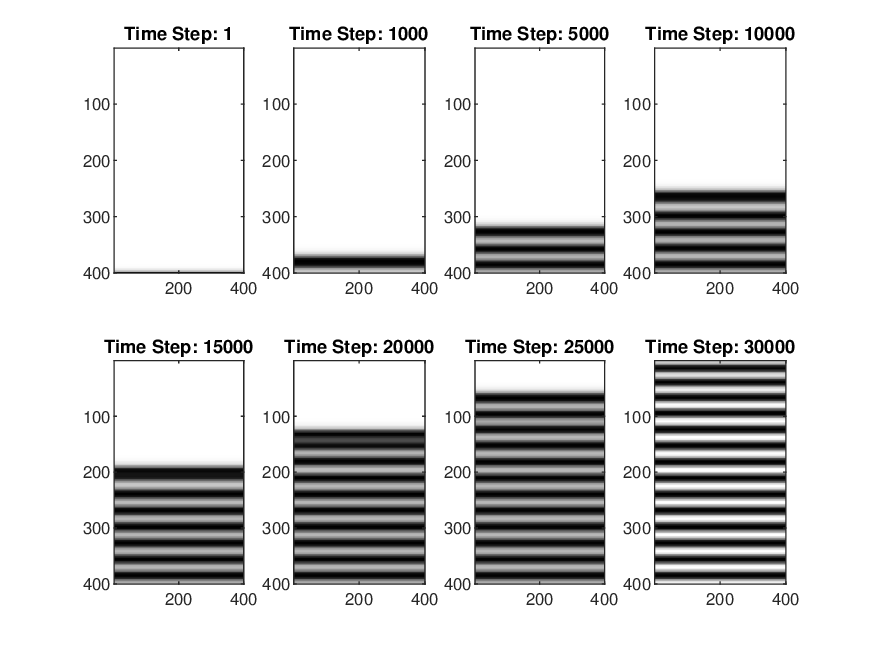}
    \caption{Simulated striped Rock Pattern at different Time Steps}
    \label{fig:simulated-striped1}
\end{figure}

\section{Metaphorical Relationship between the MATLAB Code and striped Rock Formation}

The provided MATLAB code simulates the Gray-Scott reaction-diffusion system and can be metaphorically related to the formation process of striped rock as follows:

\begin{enumerate}
    \item \textbf{Percolation of Acidic Hydrothermal Fluid}:
        \begin{itemize}
            \item In the Gray-Scott model, the grid represents the geological substrate, and the variables $u$ and $v$ represent different components of the fluid percolating through the rock.
            \item The initial conditions for $u$ and $v$ represent the introduction of the Fe2+-bearing acidic hydrothermal fluid into the rock substrate.
        \end{itemize}
        
    \item \textbf{Neutralization Reactions and Fe-Precipitation}:
        \begin{itemize}
            \item As the fluid represented by $u$ and $v$ diffuses into the grid, it encounters regions with different concentrations of chemical components, analogous to encountering primary carbonate minerals in the rock.
            \item The reaction terms in the Gray-Scott equations represent chemical reactions, including neutralization reactions between the acidic fluid and carbonate minerals, leading to the precipitation of Fe-oxyhydroxide ($u$) at the reaction front.
        \end{itemize}
        
    \item \textbf{Step-wise Progression of Fe-Precipitation}:
        \begin{itemize}
            \item Continuous diffusion of fluid components leads to the progression of chemical reactions and the formation of successive bands of precipitated minerals, similar to the step-wise progression of Fe-precipitation forming the characteristic rhythmic banding pattern in striped rock.
        \end{itemize}
        
    \item \textbf{Hydrothermal Alteration of Primary Minerals}:
        \begin{itemize}
            \item The continuous alteration of chemical concentrations in the Gray-Scott model reflects the hydrothermal alteration of primary minerals in the rock substrate due to the acidic fluid.
            \item Differences in temperature and pH of the fluid, as well as variations in mineral composition, contribute to the differentiation between different types of striped rock bands.
        \end{itemize}
        
    \item \textbf{Oxidation and Transformation to Hematite}:
        \begin{itemize}
            \item Over time, chemical reactions within the Gray-Scott model lead to the transformation of chemical components, akin to the oxidation of Fe-oxyhydroxide to hematite in the formation of striped rock.
        \end{itemize}
\end{enumerate}

The visualization aspects of the code help illustrate the evolution of the chemical concentrations ($u$) over time, providing insights into the development of patterns akin to the banding patterns observed in striped rock.

\section{Natural striped Rock Pattern}

These striped patterns result from complex geological processes that occur over extended periods. The formation of striped rock patterns in nature can be influenced by various factors, including mineral composition, sedimentation, and tectonic activity.

\begin{figure}[h!]
    \centering
\includegraphics[width=0.75\textwidth, height=0.5\textheight]{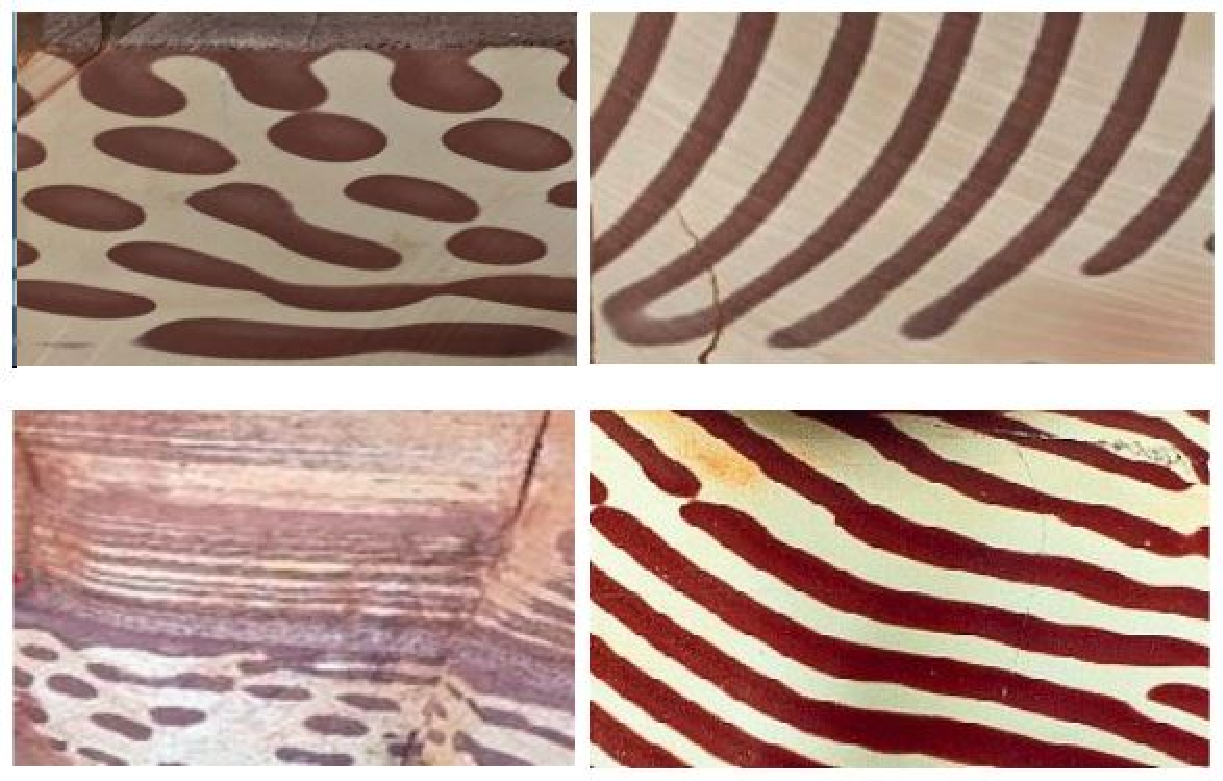}
   \caption{striped Rock Pattern\cite{coward2023}}
    \label{fig:striped_rock}
\end{figure}

\subsection{Analysis and Insights}
Comparing the simulated striped rock patterns with natural patterns allows us to draw valuable conclusions. The simulated patterns closely resemble natural striped rock formations, it suggests that the Gray-Scott Reaction-Diffusion model captures essential processes responsible for these geological features. On the other hand, different additional factors need to be considered in the model  parameters should be adjusted to better mimic real-world scenarios. This comparative analysis provides an opportunity to understand the strengths and limitations of our numerical simulation and how well it can replicate the fascinating striped rock patterns observed in geological systems. In the following sections, we will present the images of the simulated striped rock pattern and showcase the comparison with natural striped rock patterns for a more comprehensive analysis.
\begin{figure}[h!]
    \centering
    \begin{subfigure}[b]{0.75\textwidth}
        \centering
        \includegraphics[width=1\textwidth, height=0.3\textheight]{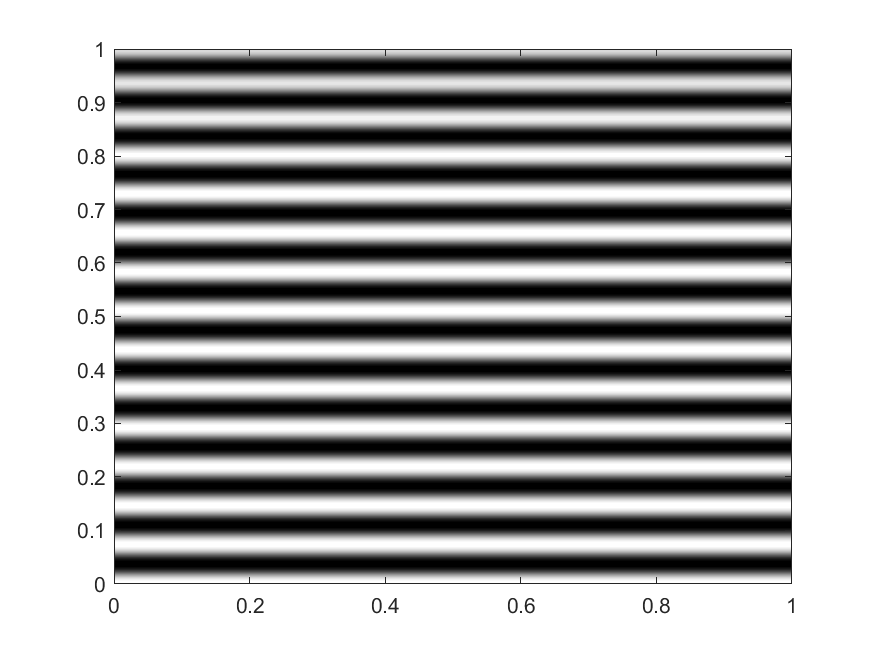}
        \caption{Simulated striped Rock Pattern}
        \label{fig:simulated}
    \end{subfigure}
    \hspace{2cm} % Adjust the horizontal space between the figures as needed
    \begin{subfigure}[b]{0.75\textwidth}
        \centering
        \includegraphics[width=0.77\textwidth, height=0.25\textheight]{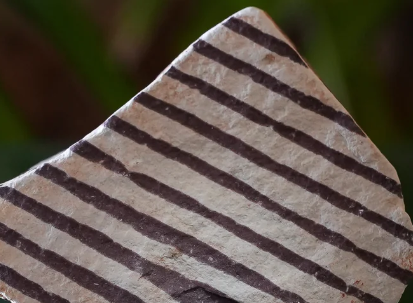}
        \caption{Natural striped Rock Pattern}
        \label{fig:natural}
    \end{subfigure}
    \caption{Comparison of Simulated and Natural striped Rock Patterns}
    \label{fig:comparison}
\end{figure}

\subsection{Conclusion}
The striped Rock Pattern is an intriguing geological marvel that captures the imagination of both scientists and enthusiasts. Its unique appearance and geological importance make it a subject of keen interest and study. By combining geological knowledge with the Gray-Scott reaction-diffusion model, we aim to uncover the intricate processes responsible for forming these distinctive patterns in rocks. Detailed mineralogical data, particularly concerning the distribution of minerals like hematite, plays a crucial role in refining and validating our model's predictions. This interdisciplinary approach bridges the gap between observed geological phenomena and microscopic processes, providing fresh insights into the enigmatic striped patterns found in rocks. Ongoing numerical simulations and sensitivity analyses will further enhance our understanding of the specific conditions that lead to the emergence of these captivating patterns.

%\section*{References}

\end{document}